\documentclass{article}
\usepackage{spconf,amsmath,graphicx}
\usepackage{stfloats}
\usepackage{enumitem}
\setlist{nosep, leftmargin=14pt}

\usepackage{mwe} 
\usepackage{svg}
\usepackage{pdfpages}

\title{CMU-NET: A STRONG CONVMIXER-BASED MEDICAL ULTRASOUND IMAGE SEGMENTATION NETWORK}
\name{Fenghe Tang$^{1}$, Lingtao Wang$^1$, Chunping Ning$^2$, Min Xian$^3$, and Jianrui Ding$^{1\star}$}
\address{$^1$School of Computer Science and Technology, Harbin Institute of Technology, Harbin, China. \\$^2$Ultrasound Department, The Affiliated Hospital of Qingdao University, Qingdao, China. \\$^3$Department of Computer Science, University of Idaho, Idaho Falls, ID 83401, USA}

\begin{document}
%
\maketitle
\begin{abstract}
U-Net and its extensions have achieved great success in medical image segmentation. However, due to the inherent local characteristics of ordinary convolution operations, U-Net encoder cannot effectively extract global context information. In addition, simple skip connections cannot capture salient features. In this work, we propose a fully convolutional segmentation network (CMU-Net) which incorporates hybrid convolutions and multi-scale attention gate. The ConvMixer module extracts global context information by mixing features at distant spatial locations. Moreover, the multi-scale attention gate emphasizes valuable features and achieves efficient skip connections. We evaluate the proposed method using both breast ultrasound datasets and a thyroid ultrasound  image dataset; and CMU-Net achieves average Intersection over Union (IoU) values of 73.27\% and 84.75\%, and F1 scores of 84.16\% and 91.71\%. The code is available at https://github.com/FengheTan9/CMU-Net.
\end{abstract}
\section{Introduction}
\label{sec:intro}
Ultrasound imaging is non-invasive, non-radiative, cost effective and real-time for disease detection. It has been widely used in the detection of breast tumor, thyroid nodule, fetal, and gonadal tissue \cite{huang2018machine}. Conventional disease detection using ultrasound images depended on  manual labeling, which is laborious and time-consuming. The results were sensitive to subjective factors such as radiologists’ experience and mental state. With the emergence of deep learning approaches, automatic medical image segmentation has been rapidly developed in the field of image analysis, which can effectively overcome the above limitations. 

The segmentation of medical ultrasound images is challenging. As shown in Fig.1, most ultrasound images only contain one lesion, and binary segmentation approaches could be applied. But the sizes, shapes, and texture patterns of lesions from different cases vary greatly. In addition, ultrasound images usually have low contrast, high speckle noises, and 
\begin{figure}[t]
\begin{minipage}[b]{1.0\linewidth}
  \centering
  \centerline{\includegraphics[width=8.5cm]{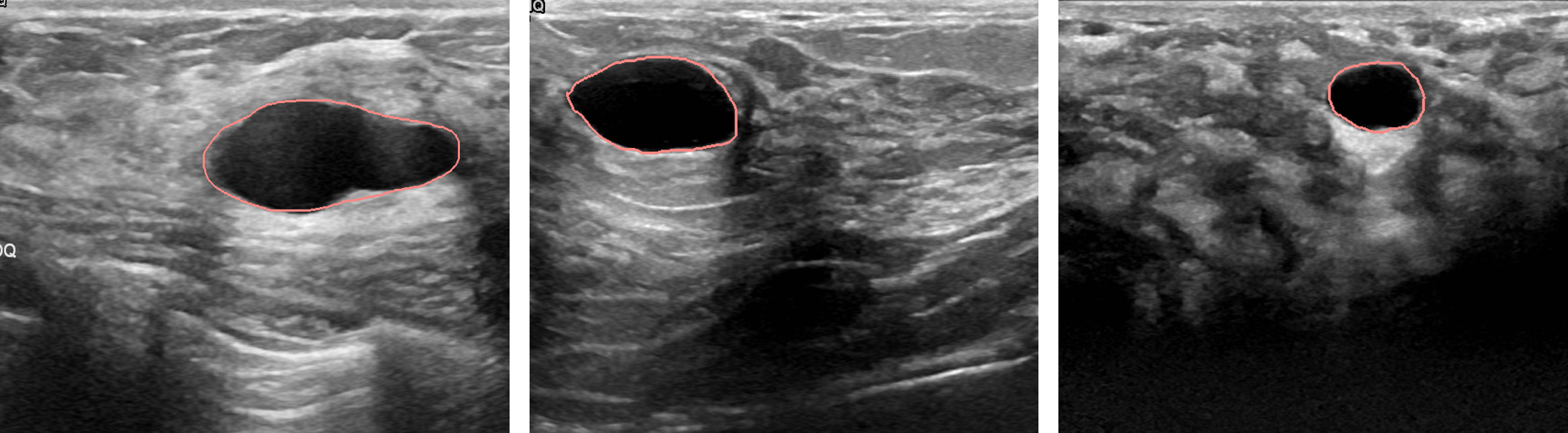}}
  \centerline{(a) Breast ultrasound image}\medskip
\end{minipage}
\vspace{-0.3cm}
\begin{minipage}[b]{1.0\linewidth}
  \centering
  \centerline{\includegraphics[width=8.5cm]{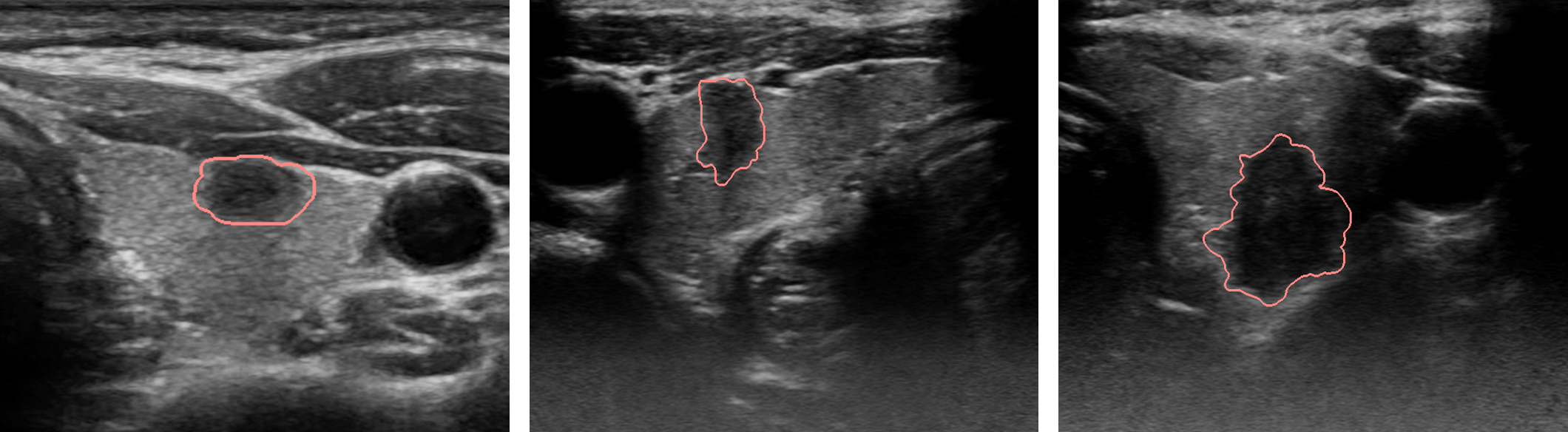}}
  \centerline{(b) Thyroid ultrasound image}\medskip
\end{minipage}
\caption{ Examples of ultrasound image segmentation. The pink contours denote lesion boundaries.
}
\label{fig:res}
\vspace{-0.2cm}
\end{figure}
shadow artifacts, and conventional segmentation approaches used to perform poorly.

U-Net \cite{ronneberger2015u} has an encoder-decoder based segmentation architecture. It can effectively fit scarce medical image data. In recent years, many medical segmentation networks based on U-Net have been proposed, such as U-Net++ \cite{zhou2018unet++}, Attention U-Net \cite{oktay2018attention}, Unet3+ \cite{huang2020unet}, and UNeXt \cite{valanarasu2022unext}. Due to the locality of ordinary convolution operations in U-Net, a number of networks based on Transformer \cite{vaswani2017attention} have recently been applied to medical image segmentation tasks \cite{chen2021transunet,valanarasu2021medical,wang2021transbts} to extract global information of images. TransUnet \cite{chen2021transunet} employed Vit \cite{dosovitskiy2020image} to obtain global context with CNN, but it required massive medical images and computing overhead.

In order to solve the limitation of ordinary convolution locality, Trockman et al. proposed the ConvMixer \cite{trockman2022patches} which used large convolutional kernels to mix remote spatial locations to obtain global context information. Compared with the Transformer, the ConvMixer is more efficiency and adapt to computer vision tasks better, and its computational overhead is less than that of the self-attention mechanism. 

Inspired by the U-shape architectural design and ConvMixer, we propose an efficient fully convolutional image segmentation network, namely CMU-Net, which contains the ConvMixer module and multiscale attention gate. The ConvMixer module is used to extract global context information. The multi-scale attention gate suppresses irrelevant features and strengthen the valuable features.

This work makes the following contributions: 1) We propose a strong fully convolution medical image segmentation network based on ConvMixer; 2) the proposed multi-scale attention gates effectively transfer knowledge using skip-connection; and 3) we successfully improve the performance of medical image segmentation using breast and thyroid ultrasound images.

\section{METHOD}
\label{sec:format}

The network architecture of the proposed CMU-Net is shown in Fig.2. The CMU-Net is divided into the encoder and decoder stages with skip-connection. In the encoder stage, high-level semantic information of medical images is extracted through ordinary convolutions, and the feature maps are input into the ConvMixer module to obtain mixed spatial and location information. In the decoder stage, the features from multi-scale attention gates are spliced with the corresponding up-sampling features to achieve accurate positioning.

\begin{figure*}[bp]

\begin{minipage}[b]{1.0\linewidth}
  \centering
  \centerline{\includegraphics[width=15cm]{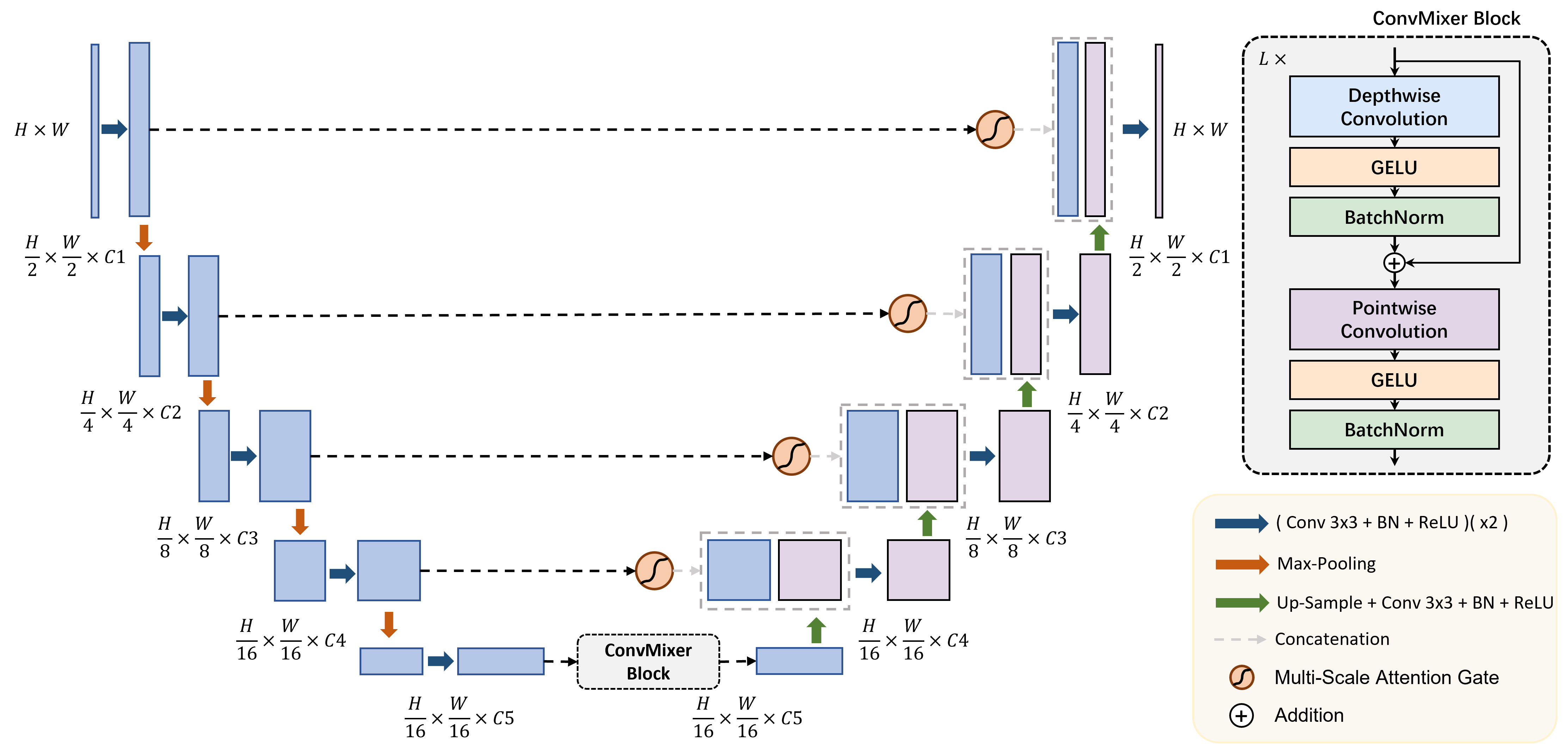}}

\end{minipage}
\caption{Overview of the proposed CMU-Net architecture. Note that the channel numbers are adopted from U-Net, i.e., C1 = 64, C2 = 128, C3 = 256, C4 = 512, and C5 = 1024.
}
\label{fig:res}
\vspace{-0.4cm}
\end{figure*}

\subsection{Encoder stage}
As shown in Fig.2, the encoder has five levels of convolutions from top to bottom. Each level consists of two ordinary convolution blocks and a down sampling operation. Specifically, each ordinary convolution block is equipped with a convolution layer, a batch normalization layer and ReLU activation. The kernel size is 3$\times$3, stride of 1 and padding of 1. The down sampling of the encoder is max pooling with window size of 2$\times$2. At the last level, the feature map is input into the ConvMixer block \cite{trockman2022patches} which is composed of  L ConvMixer layers. A single ConvMixer layer consists of depthwise convolution (i.e., kernel size k$\times$k) and pointwise convolution (i.e., kernel size 1$\times$1). The number of group channels of the depthwise convolution kernel is equal to the channels of the input feature map. Each convolution is with a GELU \cite{hendrycks2016gaussian} activation and a batch normalization, and is defined by
\begin{equation}
f_l^{\prime}=BN({\sigma}_1{\{}DepthwiseConv(f_{l-1}){\}})+f_{l-1}
\end{equation}
\begin{equation}
f_l=BN({\sigma}_1{\{}PointwiseConv(f_l^{\prime}){\}})
\end{equation}
Where $f_l$  represents the output feature map of layer $l$ in the ConvMixer block, ${\sigma}_1$ represents the GELU activation, and BN represents the batch normalization. Since the feature maps from all layers in the ConvMixer module maintain the same resolution and size, we directly up-sample the features extracted by the ConvMixer block. 

\begin{figure*}[h]
\begin{minipage}[b]{1.0\linewidth}
  \centering
  \centerline{\includegraphics[width=13cm]{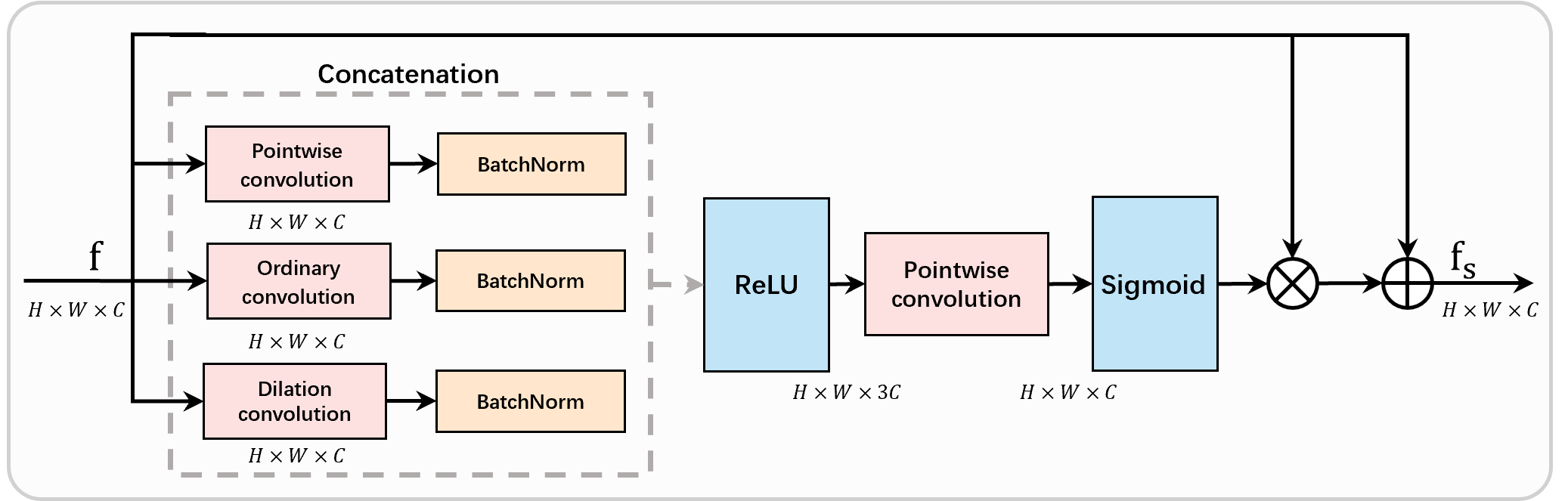}}

\end{minipage}
\caption{Multi-scale attention gate.
}
\label{fig:res}
\vspace{-0.3cm}
\end{figure*}

\subsection{Decoder stage with skip connection}
The decoder also has five modules from bottom to top. Each module is composed of two ordinary convolution blocks and an upsampling block. Specifically, the upsampling block is equipped with an upsampling layer, a convolution layer, a batch normalization layer and ReLU activation. The bilinear interpolation is utilized to upsample the feature maps. The convolution kernels have size of 3$\times$3, stride of 1 and padding of 1.

We propose the multi-scale attention gates and integrate it with the skip connections. It is used to suppress unimportant features and enhance the valuable features. Specifically, the implementation of the multi-scale attention gate is shown in Fig.3.

To select features according to different resolutions adaptively, we develop three convolutions with different receptive fields to extract features: pointwise convolution, ordinary convolution (i.e., kernel size of 3$\times$3 and stride of 1 and padding of 1) and dilated convolution (i.e., kernel size of 3$\times$3, stride of 1, padding of 2 and dilation rate of 2). Each convolution is with a batch normalization layer. The three different convolutions generate feature maps with same sizes; and we concatenate the output feature maps before a ReLU activation and vote to select valuable features by another pointwise convolution:

\begin{equation}
\begin{split}
f_{Concat}=\sigma_{2}(Concat{\{}&BN{\{}PointwiseConv(f){\}}, \\ 
&BN{\{}OrdinaryConv(f){\}},\\
&BN{\{}DilationConv(f){\}}{\}})
\end{split}
\end{equation}

\begin{equation}
f_s= f{\times} \sigma_{3}(PointwiseConv(f_{Concat}))    +f
\end{equation}
Where $f$ represents encoding features, $f_{Concat}$ is the concatenated feature, $f_s$ is the output feature from the multi-scale attention gate, and $\sigma_2$ and $\sigma_3$ denotes the ReLU and Sigmoid activation, respectively.

\begin{table*}[bp]
\vspace{-0.3cm}
\setlength\tabcolsep{8pt}
 {\caption{Results on the BUSI dataset (\%).}}\label{tab4}%
\begin{center}
\begin{tabular}{cccccc}
\hline
                       & IoU         & Recall     & Precision  & F1-value         & Accuracy   \\ \hline
U-Net{[}2{]}           & 68.49$\pm$0.18 & 80.57$\pm$2.24 & 82.52$\pm$2.34 & 80.88$\pm$0.07 & 96.74$\pm$0.08 \\
Attention U-Net{[}4{]} & 70.38$\pm$1.48  &81.44$\pm$1.67	&83.66$\pm$0.61	&82.16$\pm$0.97	&96.99$\pm$0.12 \\
U-Net++{[}3{]}         & 69.49$\pm$0.15 &	81.27$\pm$1.36	&81.87$\pm$1.07	&81.15$\pm$1.25	&96.34$\pm$0.22   \\
U-Net3+{[}5{]}         &    65.39$\pm$0.12&	77.54$\pm$2.02	&80.66$\pm$2.50	&78.22$\pm$0.07&	95.96$\pm$0.09          \\
TransUnet{[}8{]}       &  66.75$\pm$1.50	&78.65$\pm$4.32&	81.33$\pm$2.71	&79.46$\pm$1.05	&96.24$\pm$0.38  \\
UNeXt{[}6{]}           &   66.76$\pm$0.05	&77.25$\pm$1.43&	83.49$\pm$1.28	&79.72$\pm$0.12&	96.60$\pm$0.05      \\
\textbf{CMU-Net}       &   \textbf{73.27$\pm$0.43}          &   \textbf{84.26$\pm$0.54}         &      \textbf{84.81$\pm$1.32}      &     \textbf{84.16$\pm$0.47}       &           \textbf{97.33$\pm$0.14} \\ \hline
\end{tabular}
\end{center}

\end{table*}

\section{RESULTS AND DISCUSSION}
\label{sec:pagestyle}
\subsection{Datasets, evaluation and implementation details}
The Breast UltraSound Images (BUSI) \cite{al2020dataset} and private Thyroid UltraSound dataset (TUS) are utilized to evaluate the proposed approach. BUSI collected 780 breast ultrasound images, including normal, benign and malignant cases of breast cancer with their corresponding segmentation results. We only use benign and malignant images (647 images). TUS is collected from the Ultrasound Department of the Affiliated Hospital of Qingdao University. It includes 192 cases, totally 1,942 images with segmentation results from three experienced radiologists. We adopt five commonly used metrics to quantitatively evaluate the performance of different segmentation models: Intersection over Union (IoU), Recall, Precision, F1-value and Accuracy.

The loss $L$ between the predicted map $\hat{y}$ and ground truth target map $y$ is defined as a combination of binary cross entropy (BCE) and dice loss (Dice),
\begin{equation}
L=0.5BCE(\hat{y},y)+Dice(\hat{y},y)
\end{equation}

The experiments use Adam optimizer to optimize the network. The initial learning rate is set to 0.0001, and momentum is 0.9. The batch size is set to 8, and the number of epochs is 300. The two datasets are randomly split thrice, 80\% for training and 20\% for validation. In addition, we resize all images to 256$\times$256 and perform random rotation and flip for data augmentation.

\begin{table*}[h!]

\setlength\tabcolsep{8pt}
 {\caption{Results on TUS dataset (\%).}}\label{tab4}%
\begin{center}
\begin{tabular}{cccccc}
\hline
                       & IoU         & Recall     & Precision  & F1-value         & Accuracy   \\ \hline
U-Net{[}2{]}           & 83.51$\pm$0.10&	90.15$\pm$0.91&	92.00$\pm$0.83&	90.97$\pm$0.05	&99.21$\pm$0.01 \\
Attention U-Net{[}4{]} & 83.90$\pm$0.14	&90.87$\pm$0.58	&91.71$\pm$0.41	&91.21$\pm$0.08	&99.22$\pm$0.01 \\
U-Net++{[}3{]}         & 84.23$\pm$0.33	&90.59$\pm$0.35&	92.34$\pm$0.32&	91.40$\pm$0.20&	99.22$\pm$0.03   \\
U-Net3+{[}5{]}         & 83.60$\pm$0.14	&90.21$\pm$0.90	&92.02$\pm$0.78	&91.01$\pm$0.07	&99.18$\pm$0.01      \\
TransUnet{[}8{]}       & 82.75$\pm$0.25	&89.51$\pm$0.19	&91.66$\pm$0.23	&90.47$\pm$0.13	&99.13$\pm$0.02  \\
UNeXt{[}6{]}           &  81.19$\pm$0.18&	88.41$\pm$1.13&	90.86$\pm$1.17&	89.50$\pm$0.07&	99.05$\pm$0.02   \\
\textbf{CMU-Net}       &   \textbf{84.75$\pm$0.30}          &   \textbf{91.53$\pm$0.37}         &      \textbf{92.02$\pm$0.13}      &     \textbf{91.71$\pm$0.17}       &           \textbf{99.27$\pm$0.01} \\ \hline
\end{tabular}
\end{center}
\vspace{-0.4cm}
\end{table*}

\subsection{Results}
We compare CMU-Net, U-Net \cite{ronneberger2015u}, U-Net++ \cite{zhou2018unet++}, Attention U-Net \cite{oktay2018attention}, U-Net3+ \cite{huang2020unet}, TransUnet \cite{chen2021transunet} and UNeXt \cite{valanarasu2022unext}. Note that the encoder of U-Net++ is ResNet34 \cite{he2016deep}. Moreover, we conduct experiments with different ConvMixer depths and kernel sizes and finally the ConvMixer block with a depth of 7 and a kernel size of 7 can achieve the best performance.

As shown in Tables 1 and 2, CMU-Net obtains better segmentation performance than all other six approaches. Especially, on the BUSI dataset, CMU-Net achieves the highest performance and improves by 2.89\% in IoU and 2.00\% in F1 score. Further for the TUS dataset, CMU-Net improves all metrics compared with other models and achieves a better trade-off between recall and specificity. The TransUnet, which consists of hybrid CNN and Transformer structures, needs to feed a large amount of training data, performs unsatisfactorily on small datasets. On the contrary, CMU-Net demonstrates its advantages with consideration to both local and global information, and highlights features that are meaningful for the task. Some sample results are shown in Fig.4. It can be seen that CMU-Net generates more accurate lesion regions and shapes.
\renewcommand{\dblfloatpagefraction}{.9}
\begin{figure*}[bp]
\hspace{.22in}
\begin{minipage}[b]{0.05\linewidth}
  \centering
  \centerline{\includegraphics[width=1.9cm]{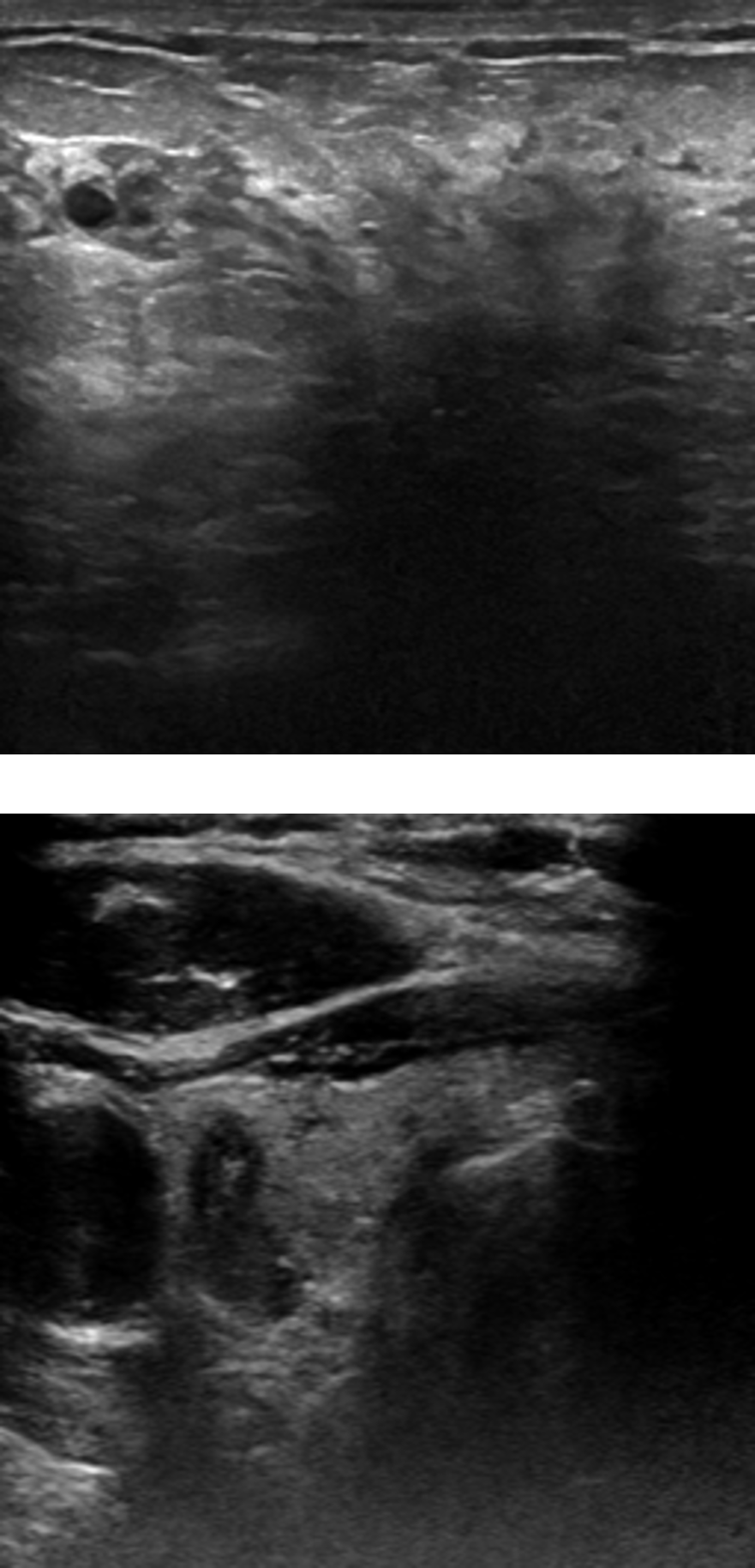}}
  \centerline{(a)}
\end{minipage}
\hspace{.35in}
\begin{minipage}[b]{0.05\linewidth}
  \centering
  \centerline{\includegraphics[width=1.9cm]{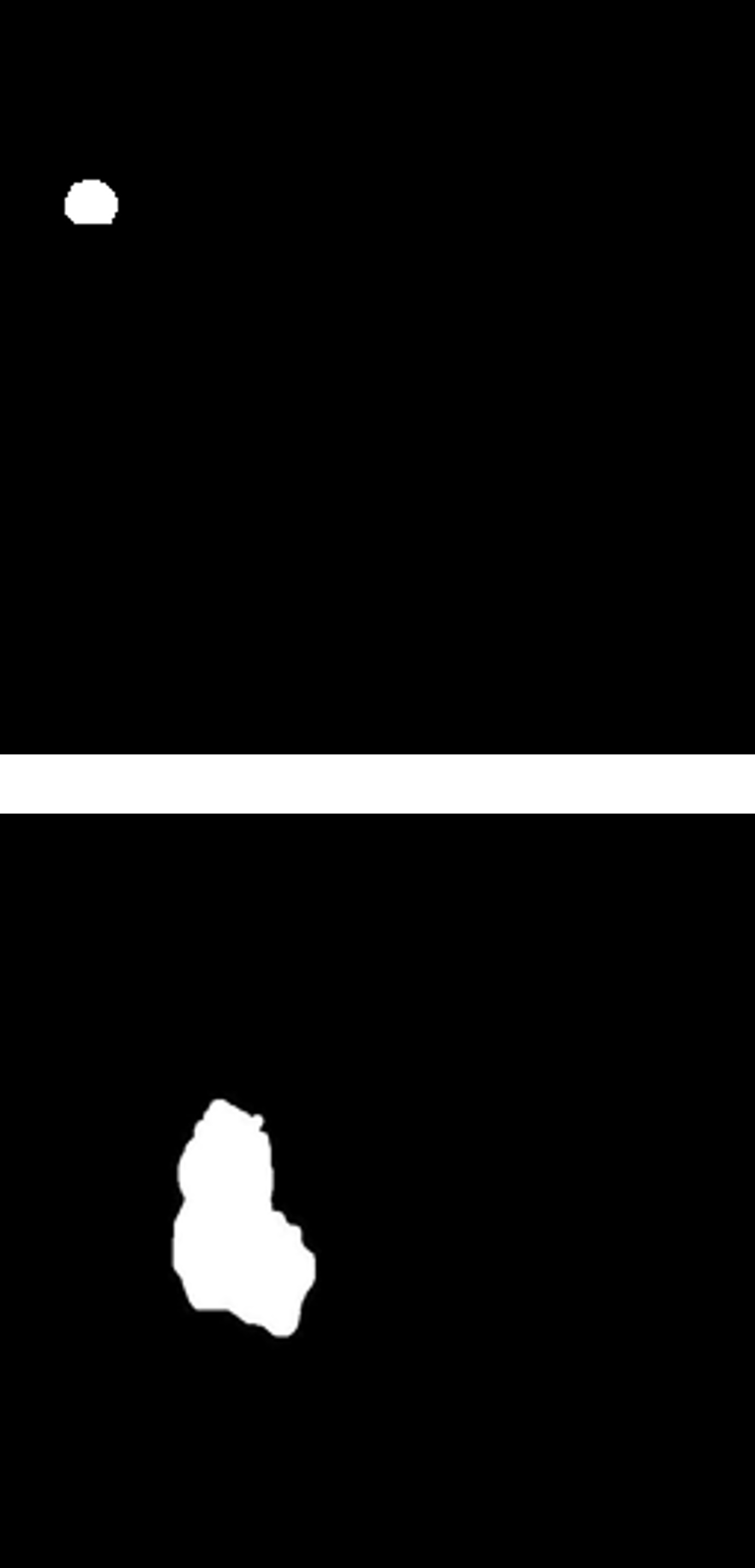}}
  \centerline{(b)}
\end{minipage}
\hspace{.35in}
\begin{minipage}[b]{0.05\linewidth}
  \centering
  \centerline{\includegraphics[width=1.9cm]{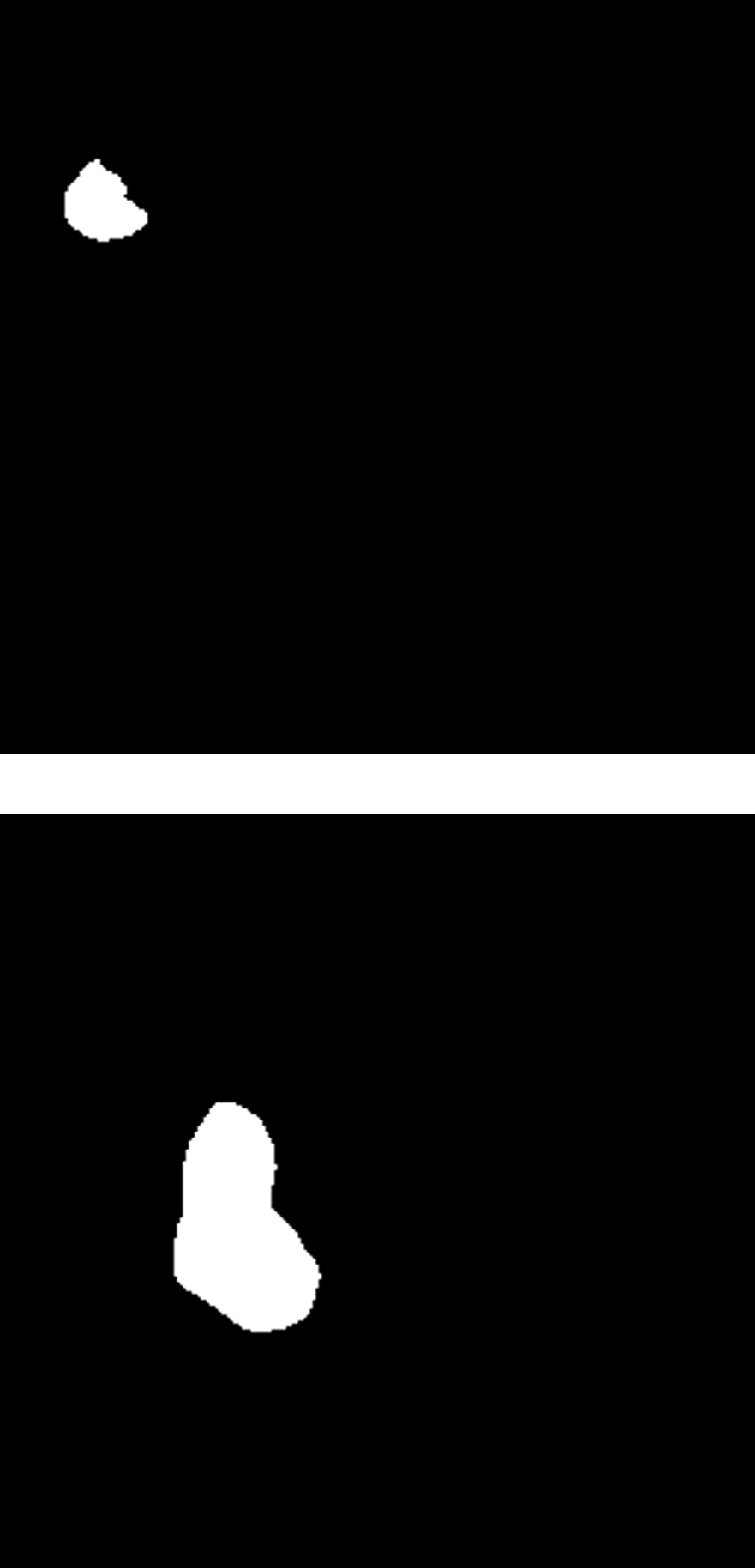}}
  \centerline{(c)}
\end{minipage}
\hspace{.35in}
\begin{minipage}[b]{0.05\linewidth}
  \centering
  \centerline{\includegraphics[width=1.9cm]{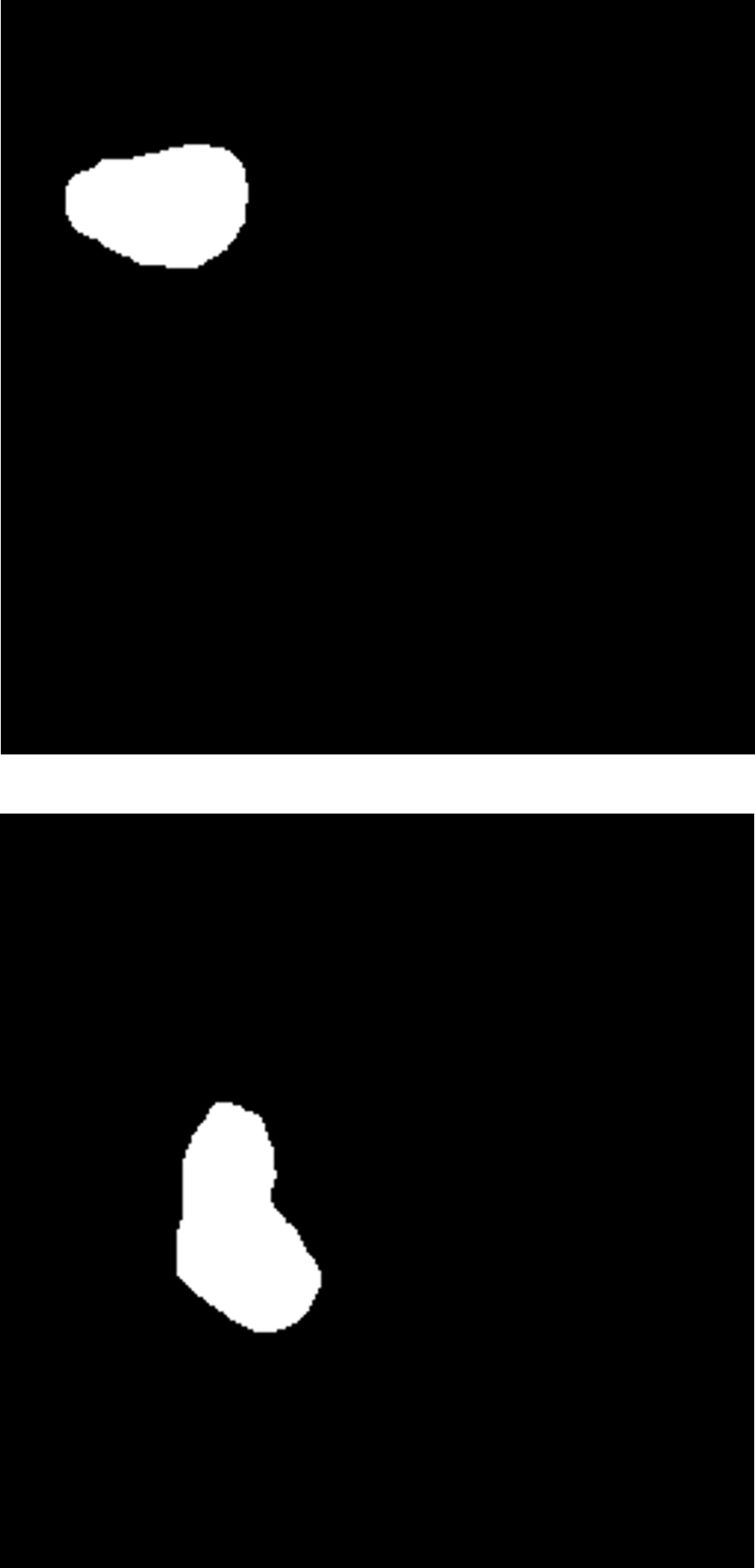}}
  \centerline{(d)}
\end{minipage}
\hspace{.35in}
\begin{minipage}[b]{0.05\linewidth}
  \centering
  \centerline{\includegraphics[width=1.9cm]{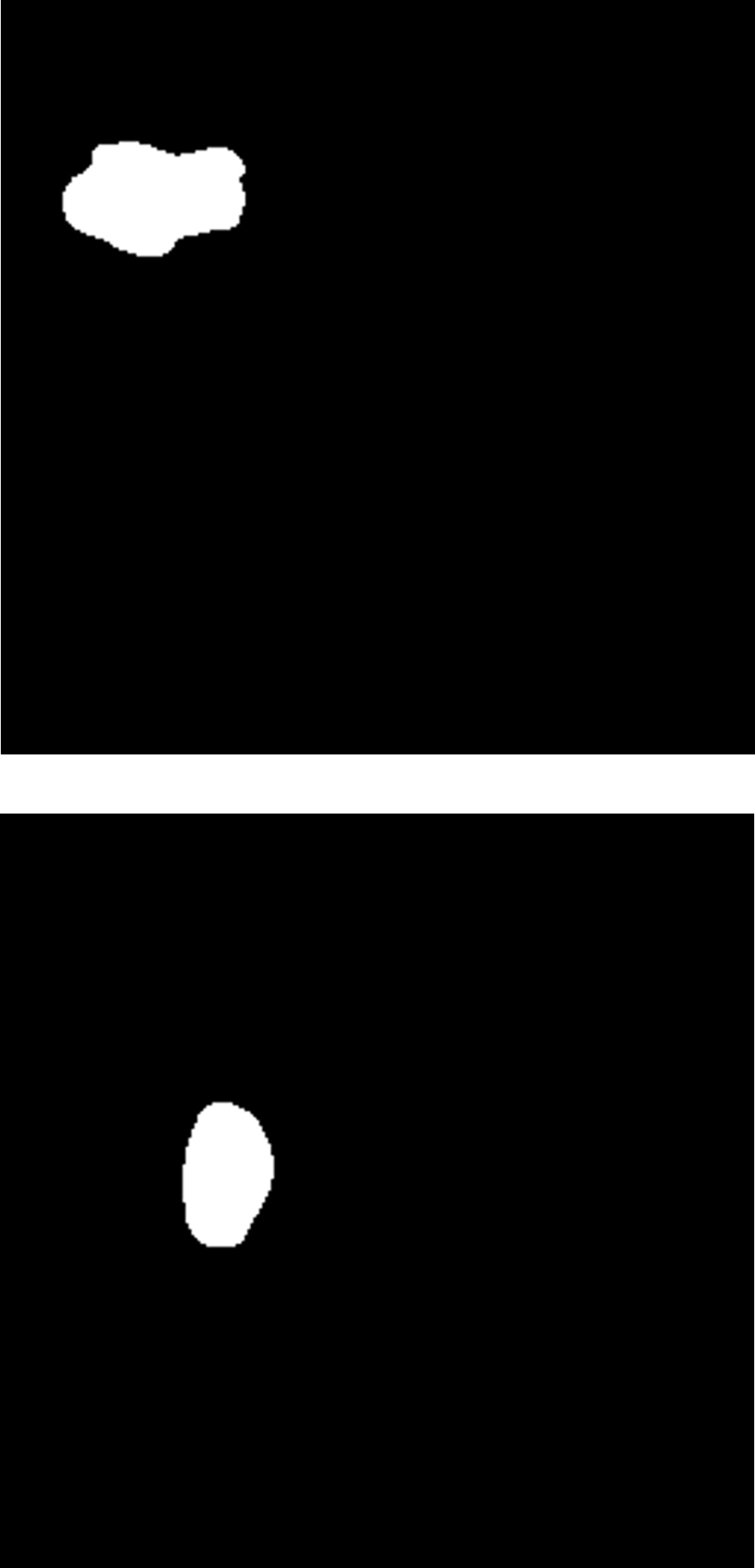}}
  \centerline{(e)}
\end{minipage}
\hspace{.35in}
\begin{minipage}[b]{0.05\linewidth}
  \centering
  \centerline{\includegraphics[width=1.9cm]{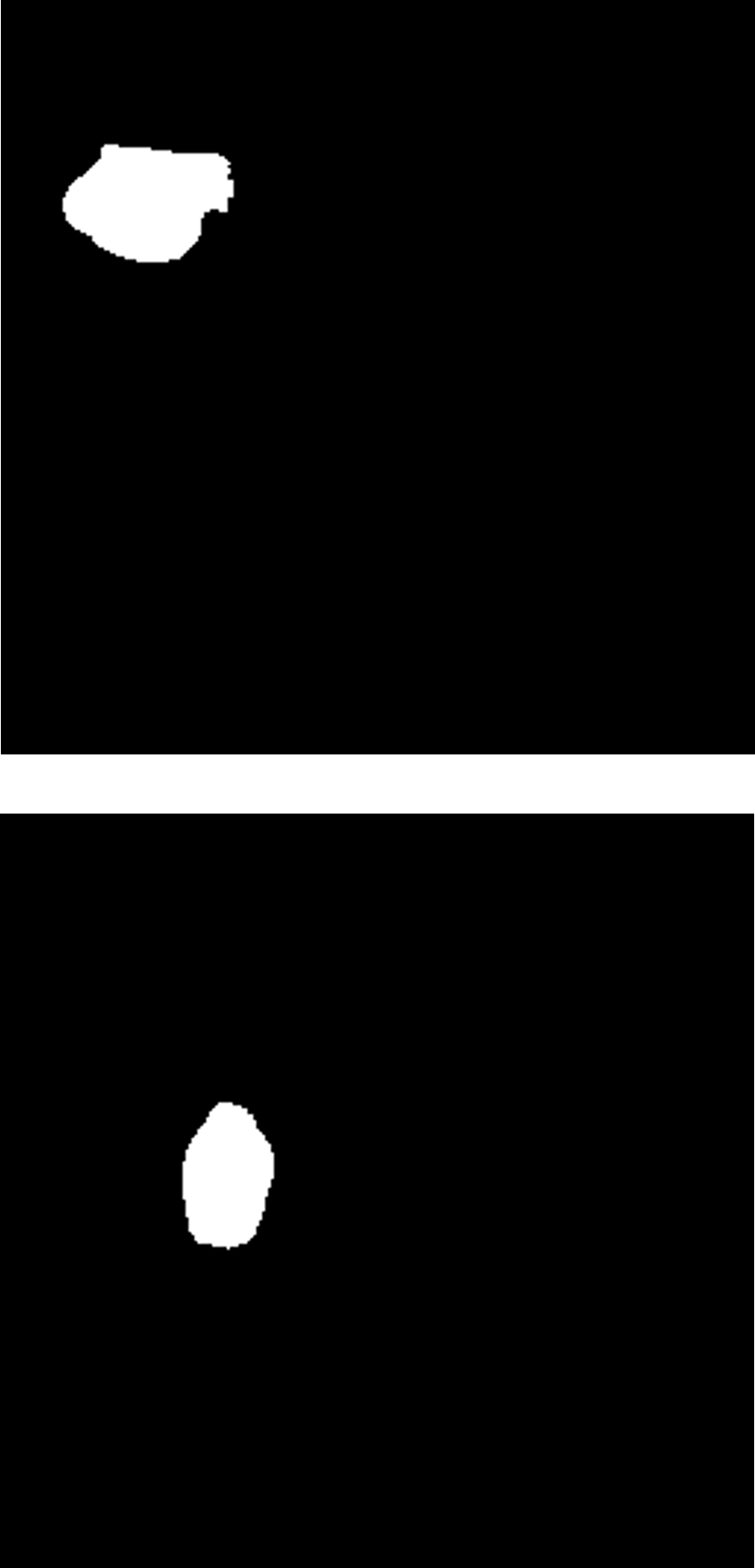}}
  \centerline{(f)}
\end{minipage}
\hspace{.35in}
\begin{minipage}[b]{0.05\linewidth}
  \centering
  \centerline{\includegraphics[width=1.9cm]{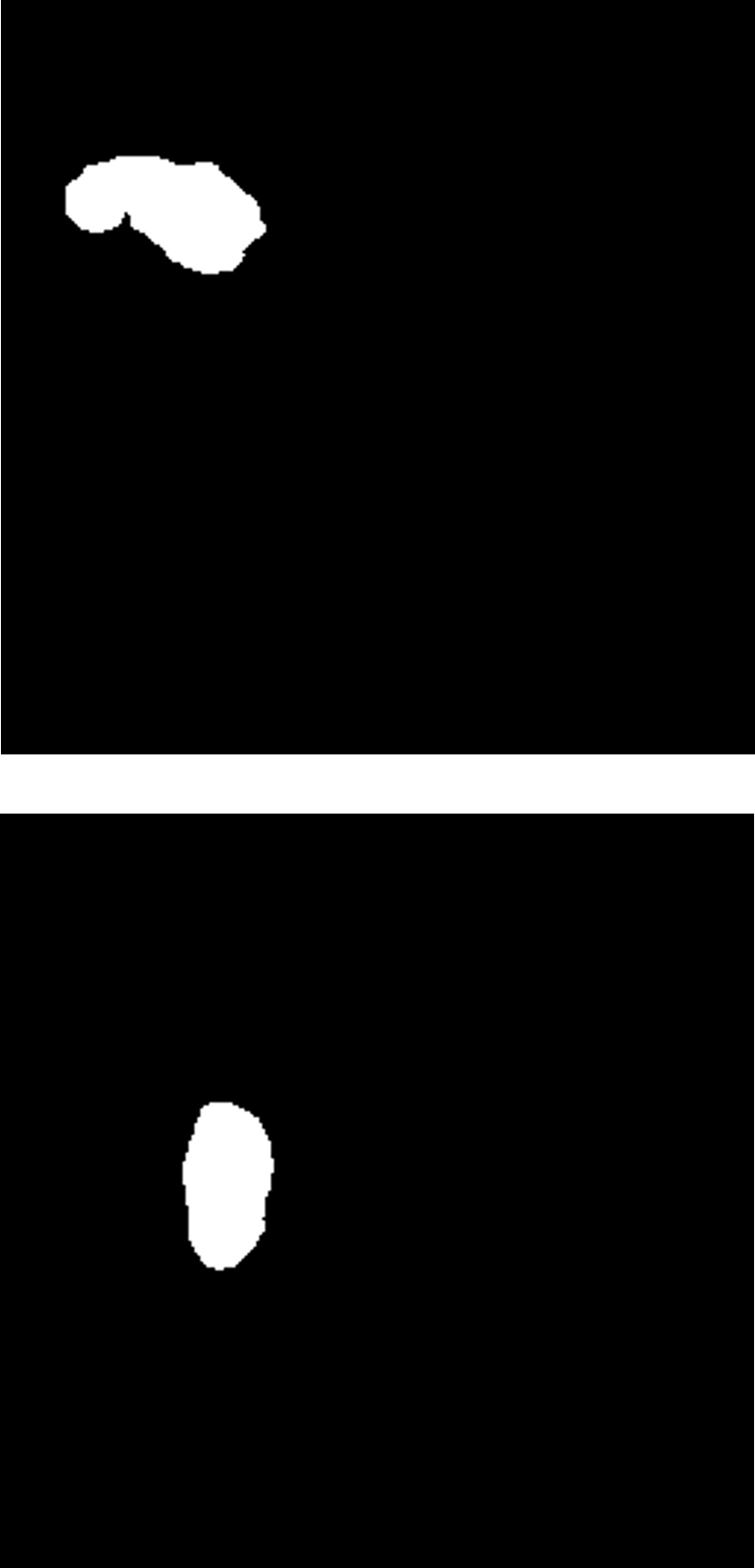}}
  \centerline{(g)}
\end{minipage}
\hspace{.35in}
\begin{minipage}[b]{0.05\linewidth}
  \centering
  \centerline{\includegraphics[width=1.9cm]{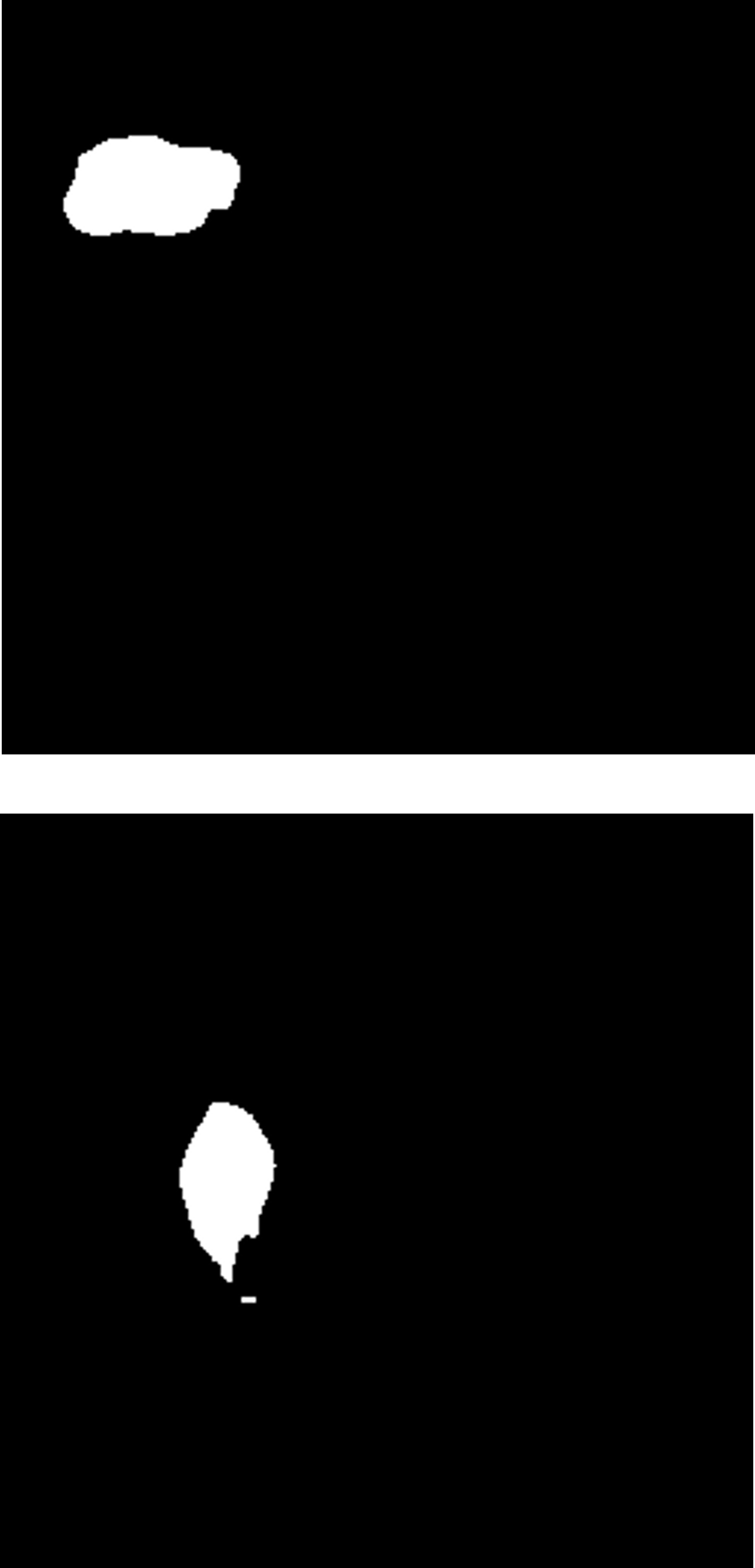}}
  \centerline{(h)}
\end{minipage}
\hspace{.35in}
\begin{minipage}[b]{0.05\linewidth}
  \centering
  \centerline{\includegraphics[width=1.9cm]{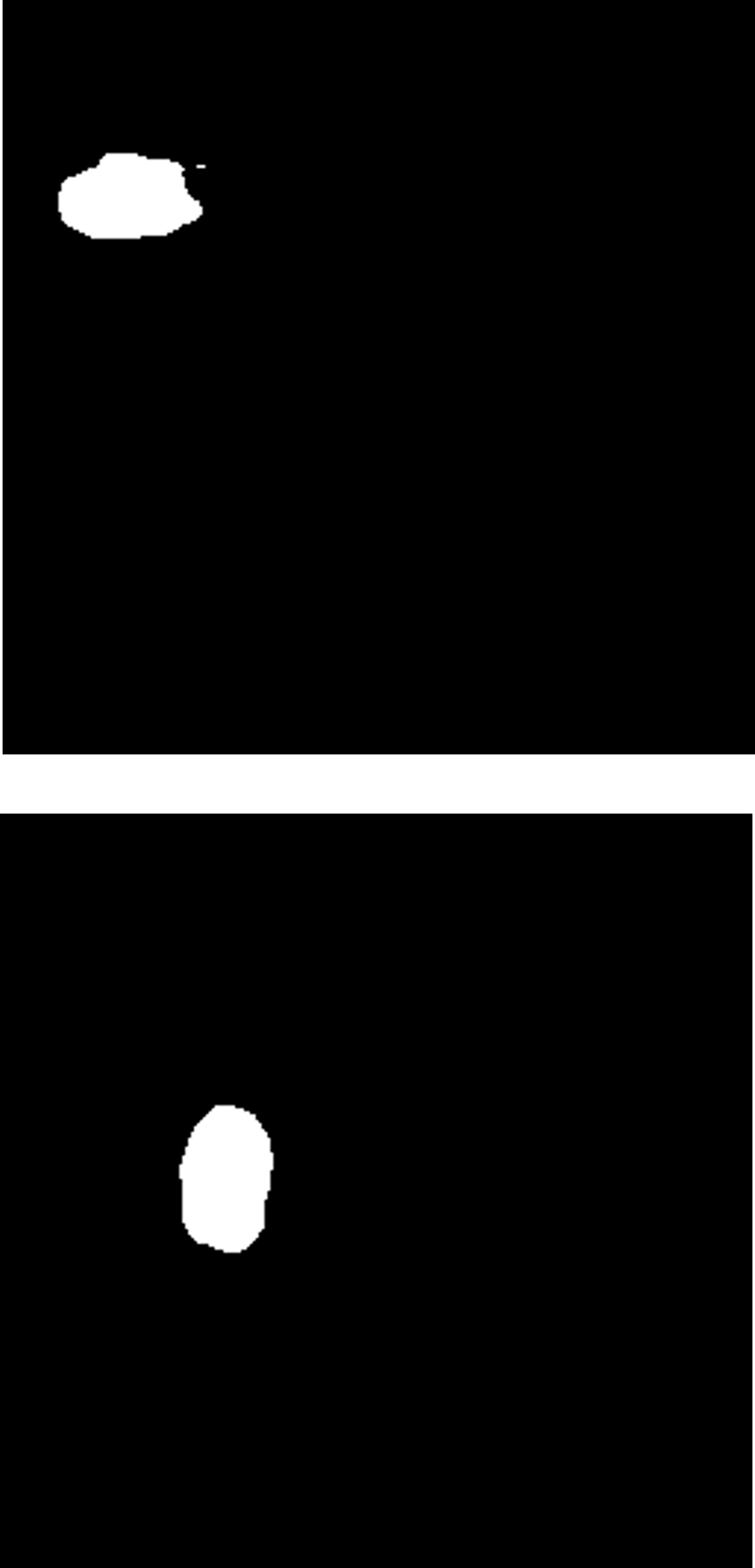}}
  \centerline{(i)}
\end{minipage}
\caption{ Segmentation result. Row 1 - BUSI dataset, Row 2 – TUS dataset. (a) Input and (b) Ground Truth. Predictions of (c) CMU-Net, (d) Attention U-Net, (e) TransUnet, (f) U-Net, (g) U-Net++ (h), U-Net3+, and (i) UNeXt.
}

\label{fig:res}
\end{figure*}
\begin{table*}[bp]
\vspace{-0.2cm}
\setlength\tabcolsep{8pt}
 {\caption{Ablation study on the BUSI dataset (\%).}}\label{tab4}%
\begin{center}
\begin{tabular}{cccccc}
\hline
                       & IoU         & F1-value         & Accuracy   \\ \hline
Original U-Net       &68.49$\pm$0.18	&80.88$\pm$0.07	&96.74$\pm$0.08  \\
U-Net + ConvMixer          &	72.36$\pm$0.37&	83.57$\pm$0.30&	97.23$\pm$0.04   \\
U-Net + ConvMixer + Multi-scale attention gate      &      73.27$\pm$0.43      &     84.16$\pm$0.47      &          97.33$\pm$0.14 \\ \hline
\end{tabular}
\end{center}
\end{table*}

Moreover, we conducted an ablation study to analyze the contribution of each module in the CMU-Net. As shown in Table 3, we can see that the performance of the original U-Net has been greatly improved after the introduction of the ConvMixer block. Next, we add multi-scale attention gates to improve the performance further, and it shows that the multi-scale attention gates can effectively magnify the affection of helpful encoder features in knowledge transfer.

\vspace{-0.2cm}
\section{CONCLUSION AND PERSPECTIVES}
\label{sec:majhead}
In this work, we propose the CMU-Net, a fully convolutional network for medical ultrasound segmentation. We introduce ConvMixer block into a U-shape network architecture to build a strong encoder for obtaining global context information, and propose a multi-scale attention gate module for emphasizing valuable features to achieve efficient skip connections. We validate CMU-Net on two ultrasound datasets, and it achieves the state-of-the-art-performance. In the future, more experiments can be carried out on CMU-Net, such as using larger convolution kernels, placing ConvMixer blocks at different encoder levels.  Further analyzing errors to improve the accuracy. Combining with the physiological and anatomical structures of the lesions to improve the interpretability of the model.

\section{Compliance with ethical standards}
\label{sec:ethics}

Informed consent was obtained from all individual participants involved in the study.

\section{Acknowledgments}
\label{sec:acknowledgments}

This work is supported, in part, by Shandong Natural Science Foundation of China; the Grant numbers is ZR2020MH290.

\bibliographystyle{IEEEbib}
\bibliography{Template_ISBI_latex}

\end{document}